\documentclass{elsart}
\newcommand{\beq}{\begin{equation}}
\newcommand{\eeq}{\end{equation}}
\newcommand{\beqar}{\begin{eqnarray}}
\newcommand{\eeqar}{\end{eqnarray}}
\newcommand{\ds}{\displaystyle}
\usepackage{epsfig}
\begin{document}
\begin{frontmatter}

\title  {
Transition to meson-dominated matter at RHIC. Consequences for
kaon flow.
         }
\author{L.V.~Bravina$^{a,b}$, L.P.~Csernai$^{c}$, Amand~Faessler$^{a}$,
C.~Fuchs$^{a}$,} 
\author{E.E.~Zabrodin$^{a,b}$
    }
\address{
$^a$Institute for Theoretical Physics, University of T\"ubingen,
    Auf der Morgenstelle 14, D-72076 T\"ubingen, Germany
   }
\address{
$^b$Institute for Nuclear Physics, Moscow State University,
    RU-119899 Moscow, Russia
   }
\address{
$^c$Department of Physics, University of Bergen,
    All\`egaten 55, N-5007 Bergen, Norway
         }


\begin{abstract}
Anisotropic flow of kaons and antikaons is studied in heavy-ion
collisions at CERN SPS and BNL RHIC energies within the microscopic
quark-gluon string model. In the midrapidity range the directed flow 
of kaons $v_1$ differs considerably from that of antikaons at SPS 
energy ($E_{lab} = 160$ AGeV), while at RHIC energy ($\sqrt{s} = 130$ 
AGeV) the excitation functions of both, kaon and antikaon, flows 
coincide within the statistical error bars. The change is attributed 
to formation of dense meson-dominated matter at RHIC, where the 
differences in interaction cross-sections of kaons and antikaons 
become unimportant. The time evolution of the kaon anisotropic flow 
is also investigated. The elliptic flow of these hadrons is found to 
develop at midrapidity at times $3 \leq t \leq 10$ fm/$c$, which is 
much larger than the nuclear passing time $t^{pass} = 0.12$ fm/$c$. 
As a function of transverse momentum the elliptic flow increases
almost linearly with rising $p_t$. It stops to rise at $p_t \geq 1.5$
GeV/$c$ reaching the saturation value $v_2^K (p_t) \approx 10\%$.
\end{abstract}

\end{frontmatter}


\section{Introduction} 
\label{intro}
The main aim of experiments on heavy-ion collisions at relativistic
and ultra-relativistic energies is to study the properties of nuclear 
matter under extreme conditions and to search for distinct signals
from the deconfined quark-gluon plasma (QGP) (see \cite{QM01} and
references therein). The transverse collective flow of particles is 
at present one of the most intensively studied characteristics of
heavy-ion collisions \cite{ReRi97,HWW99,Olli98}, because the flow 
is directly linked to the equation of state (EOS) of the system.
If even a small amount of the QGP is formed in the course of the
collision, it would lead to a reduction of pressure
\cite{Amprl91,Brplb94,CsRo99} and a softening of the EOS
\cite{HuSh95,Ri96} that can be detected experimentally. 
To study the properties of transverse particle flow the method of
Fourier series expansion \cite{VoZh96,PoVo98} has been proved to
be very useful:
\beq
\ds
E \frac{d^3 N}{d^3 p} = \frac{d^2 N}{2 \pi p_t dp_t dy} \left[ 
1 + 2 \sum_{n=1}^{\infty} v_n \cos(n\phi) \right] .
\label{eq1}
\eeq
Here $p_t$, $y$, and $\phi$ are the transverse momentum, rapidity,
and the azimuthal angle of a particle, respectively. 
The unity in square brackets represents the isotropic radial 
flow, while the other terms are refer to anisotropic flow. The 
first Fourier coefficient in Eq.~(\ref{eq1}) 
$v_1 = \langle \cos{\phi} \rangle$ is called directed flow. It 
represents the averaged ratio of the particle momentum along the
impact parameter axis to the transverse momentum, $v_1 = \langle
p_x / p_t \rangle$. The second Fourier coefficient $v_2 = \langle
\cos{(2 \phi)} \rangle$ is called elliptic flow. It characterizes
the eccentricity of the ellipsoid of the particle azimuthal 
distribution, $v_2 =\langle (p_x / p_t)^2 - (p_y/p_t)^2 \rangle $.
Methods proposed for measurement of azimuthal anisotropies in
heavy ion collisions can be found in Refs.~\cite{DaOd85,DBO01,Bor02}.

The anisotropic flow is a function of rapidity, transverse
momentum, and the impact parameter of an event $b$, i.e.
$v_n \equiv v_n(x_j)$, where $\{ x_{j=1,2,3} \} \equiv
\{ y, p_t, b \}$. Therefore, the following differential
distributions are usually applied
\beq \ds
v_n(x_i, \Delta x_{j \neq i}) = 
\int_{x_j^{(1)}}^{x_j^{(2)}} \cos(n\phi) \frac{d^3 N}{d^3 x_j} 
d^2 x_{j \neq i} \left/ \int_{x_j^{(1)}}^{x_j^{(2)}} 
\frac{d^3 N}{d^3 x_j} d^2 x_{j \neq i} \right. \ .
\label{eq2}
\eeq

The idea that the elliptic flow can carry important information
about the early stage of heavy-ion collisions has been discussed 
already in Ref.~\cite{Olli92}. This suggestion is supported by 
macroscopic hydrodynamic and microscopic transport simulations,
which show that elliptic flow saturates quite early
\cite{Sorprl97,KSH99,HeLe99,ZGK99,MoGy01}, while directed flow 
develops almost until the stage of final interactions 
\cite{LPX99,Brac00}. However, the directed flow of hadrons with high 
transverse momentum can be used as a probe of hot and dense phase of 
the collision \cite{trflprc01} due to the early freeze-out times of 
these particles.

In the present Letter we are studying the anisotropic flow of kaons,
$K^+$'s and $K^0$'s, and antikaons, $K^-$'s and $\overline{K}^0$'s,
produced in lead-lead (CERN SPS) and gold-gold (BNL RHIC) collisions 
at $E_{lab} = 160$ AGeV and $\sqrt{s} = 130$ AGeV, respectively.
Our investigation is inspired by several reasons: First of all,
in heavy-ion collisions at GSI SIS (1 - 2 GeV) and BNL AGS 
(2 - 11 AGeV) energies the excitation functions of the directed $K^+$ 
and $K^-$ flow are found to be different. At SIS energies the $K^+$ 
flow is zero or slightly anti-correlated with the nucleon flow 
\cite{fopi,kaos}. This effect becomes even more pronounced at AGS 
where a strong anti-correlation between the $K^+$ and the nucleon flow 
appears \cite{e895}. In both cases, this anti-correlation can be 
explained by the influence of an in-medium kaon potentials derived 
from effective chiral models \cite{KaNe86}. Kaons experience a weak 
repulsive potential as a result of superposition of a repulsive vector 
potential and an attractive scalar potential, whereas antikaons 
experience a strong attractive potential. Therefore, the directed 
$K^+$ flow at these energies is anti-correlated (antiflow) 
\cite{LKL95,PKLZ00} while the  $K^-$ flow is expected 
to be similar to the nucleon flow (normal flow) \cite{SLK99}.

Does it mean that at energies of SPS and higher, where the mean fields
do not play a dominant role, the anisotropic kaon flow would be similar
to that of antikaons? Also, multiple particle production at high
energies is usually described in the Monte Carlo microscopic models
via the formation and break-up of strings stretching between 
quarks and diquarks (or antiquarks). Kaons, that can be formed on
the so-called leading quarks ($u$ and $d$ quarks belonging to 
primary hadrons), are $K^+\ (u \bar{s})$ and $K^0\ (d \bar{s})$
only, but not $K^-\ (\bar{u} s)$ and $\overline{K}^0 (\bar{d} s)$.   
This may affect the flow of $K$'s, because leading hadrons are 
carrying larger transverse momenta.
Can we see the difference between the coefficients $v_n^{K^+ + K^0}$
and $v_n^{K^- + \overline{K}^0}$ in different rapidity intervals?
Finally, $K^-$'s and $\overline{K}^0$'s can be absorbed via the 
channels such as $K^- + p \rightarrow \Lambda + \pi^+$, etc.,
whereas there are no analogous reactions for $K^+$'s and $K^0$'s.
How important is this reaction asymmetry at SPS and, especially, at 
RHIC energies, where the matter is expected to be meson-dominated?

To answer these questions we employed the microscopic cascade
quark-gluon string model (QGSM) \cite{Ka82,qgsm,strange,hard}. 
Description of the model is given in Sec.~\ref{model}. Rapidity and 
transverse momentum dependences of directed (Sec.~\ref{dir_flow}) 
and elliptic (Sec.~\ref{ell_flow}) flow of kaons and antikaons are 
calculated in Pb+Pb and Au+Au collisions at SPS and RHIC energies, 
respectively. Time development of the both flows is studied as well. 
Finally, conclusions are drawn in Sec.~\ref{concl}.

\section{The model}
\label{model}
Similar to the dual parton model (DPM) \cite{dpm}, VENUS \cite{venus}, 
and the string fusion model (SFM) \cite{sfm}, QGSM is based on 
Gribov-Regge theory (GRT) \cite{GRT} accomplished by a string 
phenomenology of particle production in inelastic hadron-hadron 
({\it hh\/}) collisions. To describe hadron-nucleus and 
nucleus-nucleus collisions the cascade procedure of multiple 
secondary interactions of hadrons was implemented. The model 
incorporates the string fragmentation, formation of resonances, and
rescattering of hadrons, but simplifies the nuclear effects
neglecting, e.g., the mean fields or evaporation from spectators.
As independent degrees of freedom QGSM includes octet and decuplet
baryons, octet and nonet vector and pseudoscalar mesons, and their
antiparticles. The momenta and positions of nucleons inside the
nuclei are generated in accordance with the Fermi momentum
distribution and the Woods-Saxon density distribution, respectively.
Two particles can interact inelastically with the cross section 
$\sigma_{inel}(s)$ or elastically with the cross section 
$\sigma_{el}(s)$ with the "black disk" probability $\ds P(s,b) = 
\theta \left(\sigma(s)/\pi - b^2 \right)$. Here $\sigma(s) = 
\sigma_{inel}(s) + \sigma_{el}(s)$ is the total cross section, $b$
is the impact parameter, and $\theta$ is the step function.
Pauli blocking of occupied final states is taken into account.

Several different subprocesses are responsible for the string 
formation in inelastic collisions. A single string can be produced as 
a result of the valence (di)quark annihilation. This subprocess is
important in {\it hh\/} interactions at low energies since the
annihilation cross section drops as $s^{-1/2}$. The baryon-antibaryon
annihilation in QGSM is modeled via the string junction annihilation
with the creation of three strings. In other subprocesses two or more 
strings can be produced as a result of the colour exchange mechanism,
while in single or double diffractive scatterings one or two strings
are formed due to momentum transfer. The Pomeron, which is a pole 
with an intercept $\alpha_P(0) > 1$ in the GRT, corresponds to the
cylinder-type diagrams. The $s$-channel discontinuities of the
diagrams, representing the exchange by $n$-Pomerons, are related to
process of $2 k\, (k\leq n)$ string production. If the contributions
of all $n$-Pomeron exchanges to the forward elastic scattering 
amplitude are known, the AGK cutting rules \cite{agk} enable one to
determine the cross sections for $2k$-strings. The hard gluon-gluon 
scattering and semi-hard processes with quark and gluon interactions 
are also incorporated in the model \cite{hard}. The inclusive 
spectra in the QGSM automatically have the correct triple-Regge limit
for Feynman variable $x \rightarrow 1$, double-Regge limit for
$x \rightarrow 0$, and satisfy to all conservation laws \cite{Ka82}. 

Because of the uncertainty
principle hadrons produced in string fragmentation can interact
further only after a certain formation time. However, hadrons
containing the valence quarks of the primary nucleons can interact
promptly with reduced interaction cross section. To calculate the
statistical weights for the different subprocesses the experimental
total, elastic, and inelastic cross sections are used. If the
experimental data are lacking, the additive quark model relations,
the one-pion exchange model, detailed balance considerations, and
isotopic invariance are employed. 

The variety of subprocesses 
included in the model, - diagrams with quark annihilation 
and rearrangement, associated with the exchange of Reggeons, 
subprocesses with colour exchange which are connected to 
(multi)Pomeron exchanges, as well as hard processes with large 
momentum transfer, - is quite rich. The contribution of each 
subprocess to the inelastic cross section depends strongly on the 
bombarding energy. This circumstance enabled the model to predict 
quantitatively the directed flow of nucleons at SPS \cite{Amprl91} 
more than six years before the first measurements \cite{na49,wa98} 
and to describe the characteristics of the elliptic flow of charged 
particles at RHIC \cite{plb01} measured quite recently 
\cite{STAR,PHOBOS,PHENIX}.
The formation of a QGP is not assumed in the present version of the 
QGSM, although one may consider the strongly interacting field of
coloured strings as a precursor of the plasma. Our strategy is to
find clear quantitative and qualitative deviations of the 
experimental results from model predictions. These deviations
can then be taken as an indication of new physics, most probably
the creation of QGP, not included in the model.

\section{Directed flow}
\label{dir_flow}
\begin{figure}[htp]
\centerline{\epsfysize=16cm \epsfbox{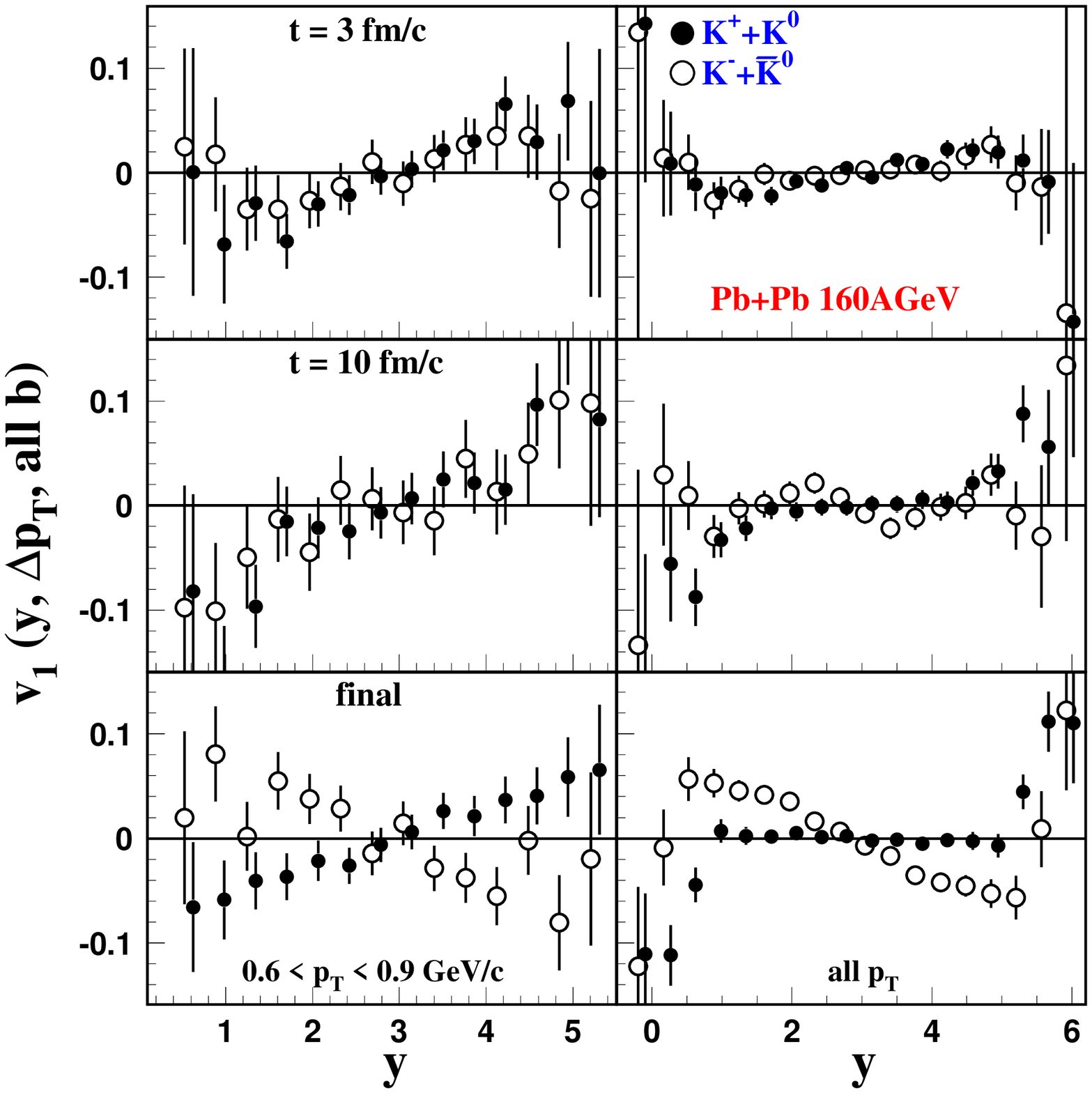}}
\vspace{0.5cm}
\caption{
The directed flow $v_1$ of kaons (solid circles) and 
antikaons (open circles) in minimum bias Pb+Pb collisions at SPS
energy as a function of rapidity $y$ in high $p_t$ interval
$0.6 \leq p_t \leq 0.9$ GeV/$c$ (left panels) and for all transverse 
momenta (right panels) at times $t = 3$ fm/$c$ (upper row), 10 fm/$c$ 
(middle row), and final (bottom row).
}
\label{fig1}
\end{figure}

Time evolution of directed flow of kaons and antikaons in minimum
bias Pb+Pb collisions at 160AGeV is presented in Fig.~\ref{fig1}.
Here the coefficient $v_1^K(y)$ is calculated in different
transverse momentum intervals at early, $t=3$ fm/$c$ and $t = 10$ 
fm/$c$, and the final stage of the reaction, $t \geq 60$ fm/c. To 
avoid ambiguities, all resonances in the scenario with early 
freeze-out were allowed to decay according to their branching ratios. 
The flow evolution is seen quite distinctly. At early stages of the 
collision directed flow of both kaons and antikaons is oriented in 
the direction of normal flow similar to that of nucleons 
\cite{trflprc01}.  Within the error bars there is no differences 
between $(K^+ + K^0)$'s and $(K^- + \overline{K}^0)$'s. 
At this stage the matter is quite dense, mean free paths of particles 
are short, and similarities in kaon production and rescattering 
dominate over inequalities caused by different interaction 
cross-sections. It is worth mentioning that the directed flow of $K$'s 
and $\overline{K}$'s is already sizeable at $t = 3$ fm/$c$. This can be 
explained by a kick-off effect associated with the early stage of the
collision, when the nuclei pass through each other. Later on the 
system becomes more dilute. For both  kaons and antikaons the directed
flow experiences significant transformations as seen in 
Fig.~\ref{fig1}. Already at $t = 10$ fm/$c$ the antiflow of 
antikaons starts to built up in the midrapidity range. Note, that 
$t = 10$ fm/$c$ corresponds to the maximum of the kaon $d N / d t$ 
distribution over their last elastic or inelastic interaction 
\cite{froprc99}. Here the differences in interaction cross sections 
and possible reaction mechanisms become crucial. As was found in 
previous investigations \cite{Brplb95,flow}, particles emitted in the 
direction of dense baryon-rich matter (i.e., normal flow) will 
interact more frequently compared to those emitted in the opposite, 
i.e. antiflow, direction. This circumstance reduces the resulting 
directed flow of kaons almost to zero in a broad midrapidity range. 
Due to larger interaction cross-sections of $K^-$'s and 
$\overline{K}^0$'s with other hadrons, the directed flow of these 
particles changes the orientation from a weak normal to strong 
antiflow. Even $(K^- + \overline{K}^0)$'s with high transverse 
momentum demonstrate distinct antiflow, while the flow of 
$(K^+ + {K}^0)$'s  remains almost unchanged compared to that at 
$t = 10$ fm/$c$.

\begin{figure}[htp]
\centerline{\epsfysize=16cm \epsfbox{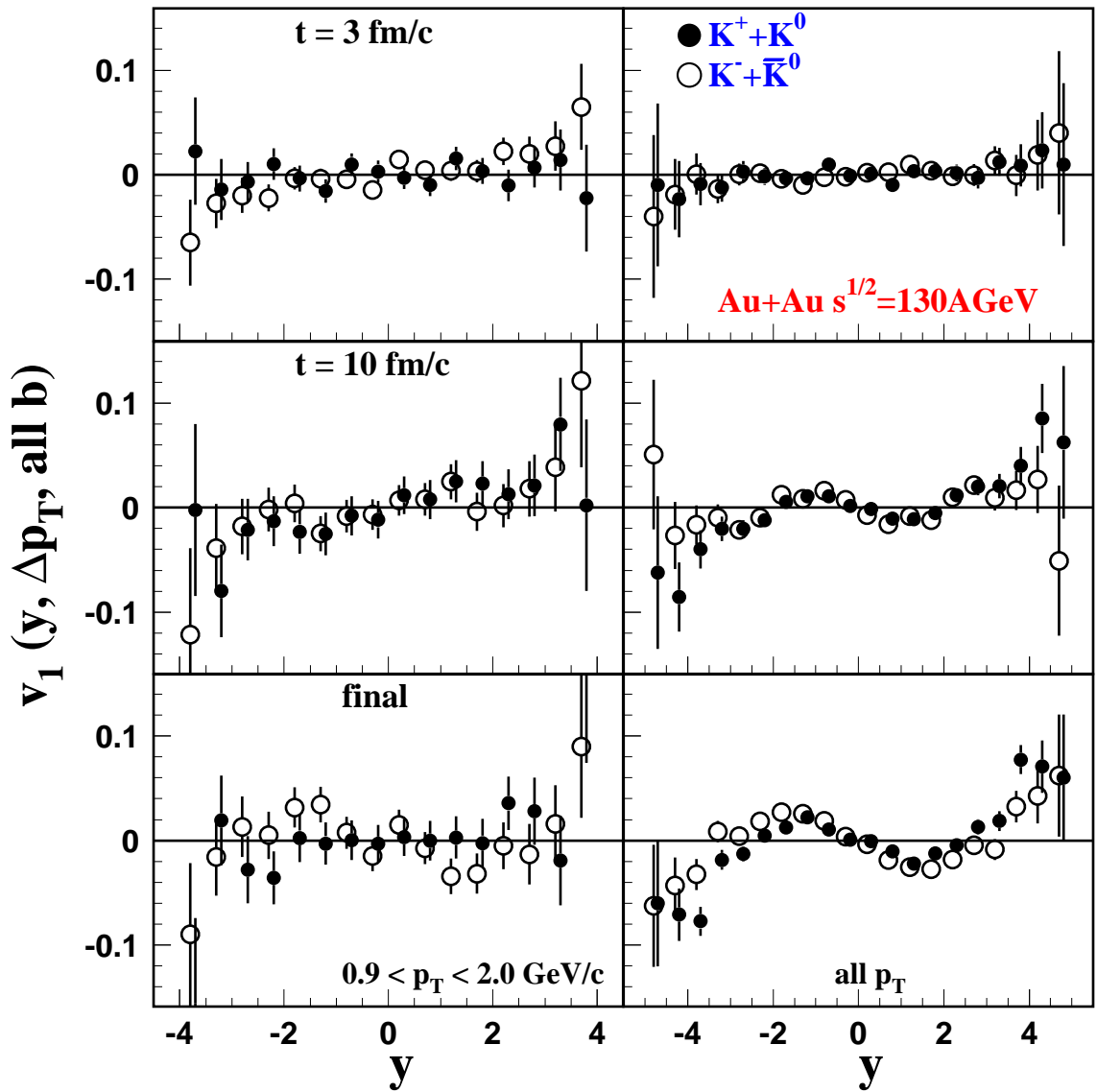}}
\vspace{0.5cm}
\caption{
The same as Fig.~1 but for minimum bias Au+Au collisions at RHIC
($s^{1/2} = 130$ AGeV). For high $p_t$-kaons transverse momentum
varies from 0.9 GeV/$c$ to 2.0 GeV/$c$.
}
\label{fig2}
\end{figure}

The directed flow of kaons and antikaons in minimum bias Au+Au 
collisions at $\sqrt{s} = 130$ AGeV is shown in Fig.~\ref{fig2}
again at early
stages, $t = 3$ fm/$c$ and $t = 10$ fm/$c$, and at the final one, 
$t \geq 100$ fm/$c$. It is interesting that at $t = 3$ fm/$c$ (i) 
the flow of $(K^+ + K^0)$'s coincides within the statistical 
errors with the $(K^- + \overline{K}^0)$ flow, and (ii) the flow is 
generally very similar to that at the SPS energy at time $t = 3$ 
fm/$c$. Except of the target and projectile fragmentation region, 
where again the flow is probably produced by the initial kick, the 
kaon flow at this early stage of gold-gold collisions at RHIC energy 
is isotropic with respect to the impact parameter axis. 
Note that the colliding nuclei have passed through 
each other already at times less than 0.12 fm/$c$. The spatial 
anisotropy in the distribution of baryonic charge seems to be 
unimportant at this stage.  At $t = 10$ fm/$c$ not only the directed 
flow of antikaons, but also that of kaons becomes antiflow-aligned at 
midrapidity. Similar behaviour has been found within the RQMD model
for the directed flow of nucleons at RHIC \cite{Sne00}, suggesting 
that the nucleon directed flow is a side effect of the elliptic flow.
The flow of produced particles, pions \cite{Sne00,BlSt00} and kaons
\cite{BlSt00}, was found to be very flat at $|y| \leq 2$, in stark
contrast to the QGSM predictions. We are awaiting the experimental 
data to resolve this problem. 

At the final stage of the reaction the excitation 
functions of both kaon and antikaon flow have a similar antiflow 
behaviour in the midrapidity range compared to the zero flow of 
$(K^+ + K^0)$'s at lower energies. However, the kaon flow at high 
transverse momenta $0.9 \leq p_t \leq 2.0$ GeV/$c$ is elongated in a 
normal direction. Its slope at the stage of thermal freeze-out is 
steeper than the slope at $t = 10$ fm/$c$, i.e., the flow of high 
$p_t$-kaons becomes stronger. This means that at RHIC energies the 
nuclear medium is rather dense, the net baryon density is small, and 
that kaon-meson (including resonances) reactions dominate over 
kaon-baryon interactions, which can significantly modify the flow 
of $K^-$'s and $\overline{K}^0$'s at SPS energies.

\section{Elliptic flow}
\label{ell_flow}
\begin{figure}[htp]
\centerline{\epsfysize=16cm \epsfbox{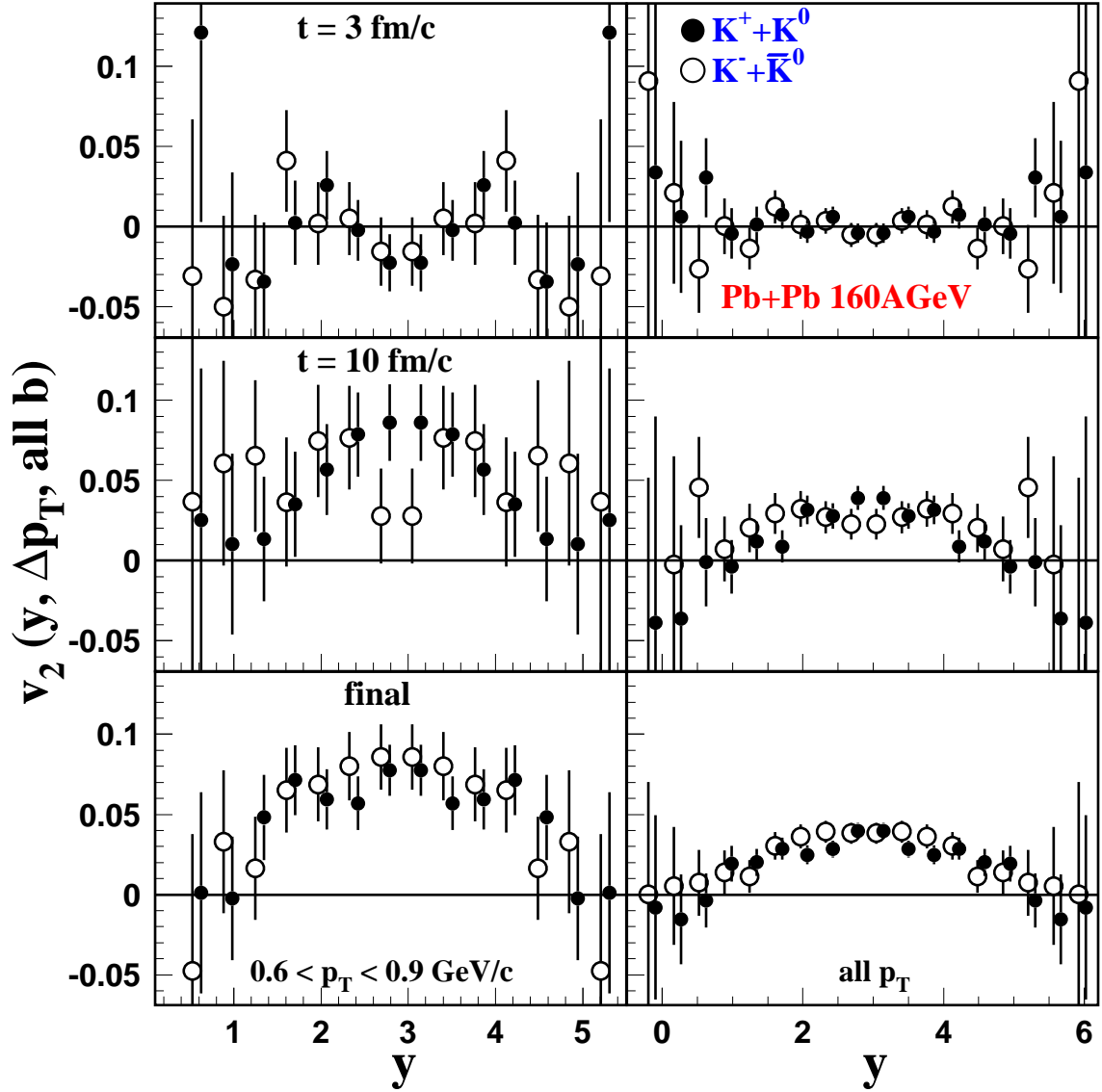}}
\vspace{0.5cm}
\caption{
The same as Fig.~1 but for the elliptic flow 
$v_2$ of kaons (solid circles) and antikaons (open circles) in
minimum bias Pb+Pb collisions at SPS.
}
\label{fig3}
\end{figure}

The study of the development of elliptic flow of kaons in minimum bias 
Pb+Pb collisions at 160 AGeV, displayed in Fig.~\ref{fig3}, reveals
that there is no particular direction of kaon emission in the
transverse plane at $t = 3$ fm/$c$. From this time on until the 
freeze-out stage one cannot distinguish between the elliptic flow of 
kaons and antikaons in the various $p_t$-intervals. The resulting 
elliptic flow of $K$'s and $\overline{K}$'s is positive, i.e., kaons 
are propagating preferentially in the reaction plane. The flow 
increases with rising transverse momentum. This behaviour appears to 
be quite similar to those obtained from the hydrodynamic calculations. 
Recall that in the hydro models the strong in-plane elliptic flow of 
hadrons is explained exclusively by the so called almond shape of the 
overlapping region of the colliding nuclei. As a consequence of this 
initial spatial anisotropy, in-plane pressure gradients should be 
stronger than their out-of-plane counterparts \cite{Olli92}. The fluid
acceleration is proportional to the pressure gradient divided by total 
energy density. During the early stage of the expansion this ratio 
remains roughly constant. When the spherical shape of the expanding 
region is restored, the pressure becomes uniform. As was argued in 
\cite{Olli92}, the typical time scale for the transverse flow is the 
transverse size of the system divided by the sound velocity in the
medium. For a semi-peripheral gold-gold or lead-lead collisions this
criterion gives us approximately 7 fm/$c$. At early times the elliptic
flow in hydrodynamics increases linearly with time. Therefore, it is
quite plausible that the elliptic flow of (anti)kaons at midrapidity 
in the QGSM calculations is formed within the time interval
$3\, {\rm fm}/c\ \leq\ t \ \leq\ 10\,{\rm fm}/c$ in the center-of-mass
(CM) system.  

\begin{figure}[htp]
\centerline{\epsfysize=16cm \epsfbox{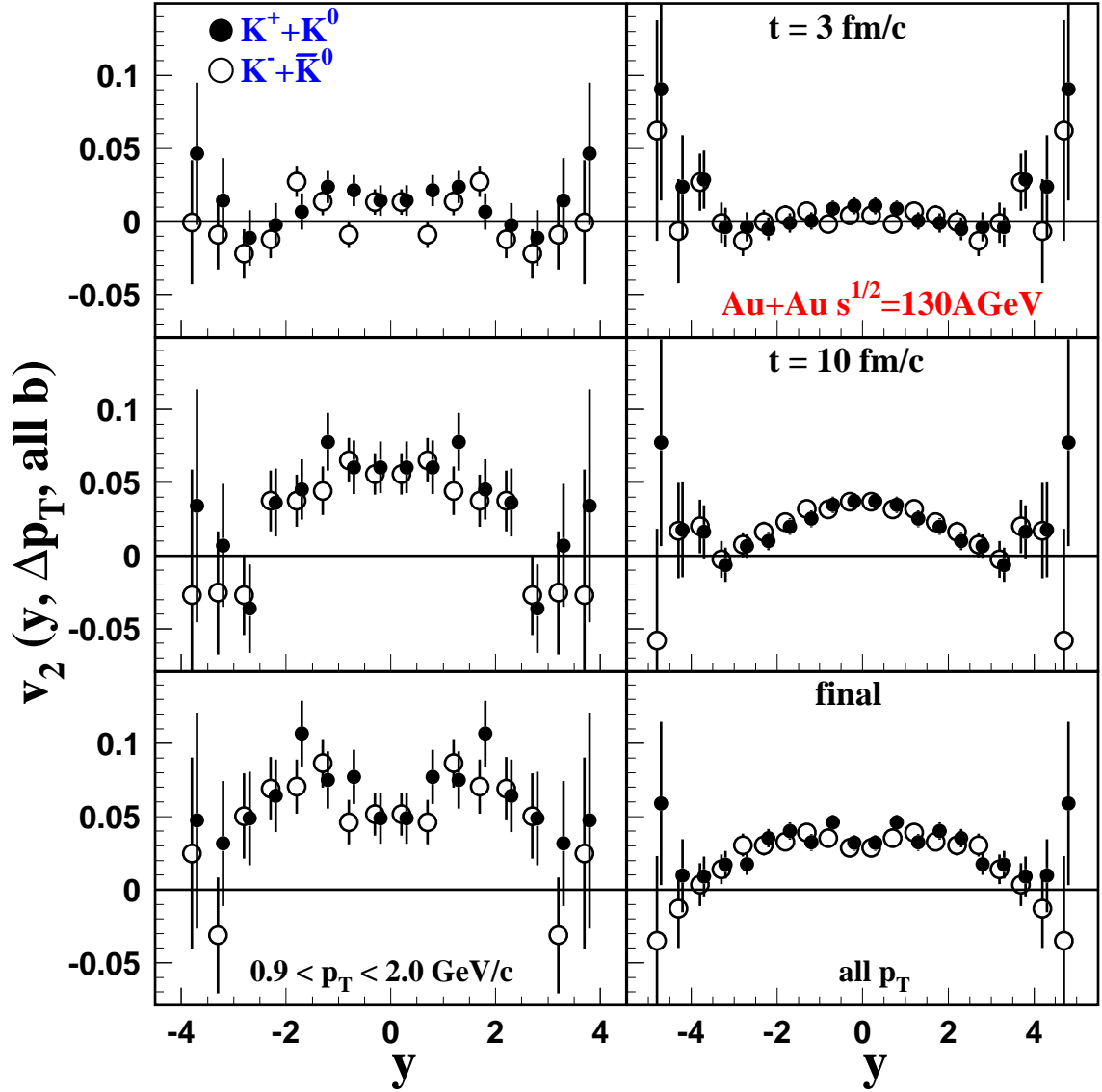}}
\vspace{0.5cm}
\caption{
The same as Fig.~2 but for the elliptic flow
$v_2$ of kaons (solid circles) and antikaons (open circles) in
minimum bias Au+Au collisions at RHIC.
}
\label{fig4}
\end{figure}

Similar features are observed in the development of the elliptic flow
of kaons and antikaons in Au+Au collisions at $\sqrt{s} = 130$ AGeV, 
as one can see in Fig.~\ref{fig4}. Their elliptic flow is close to 
zero in the midrapidity range $|y| \leq 2$ at $t = 3$ fm/$c$. Even the 
flow of high-$p_t$ particles, 
which are produced in very energetic hadronic collisions at the 
beginning of the reaction and which decouple from the system earlier 
than other particles, shows the same ``isotropic" tendency. However, 
the elliptic flow of kaons and antikaons in the midrapidity range 
$2 < y < 4$ (SPS) and $|y| \leq 1.5$ (RHIC) seems to be formed between 
3 fm/$c$ and 10 fm/$c$, i.e., the flow indeed probes the early stage 
of heavy-ion collisions.  The final elliptic flow of both, 
$(K^+ + K^0)$'s and $(K^- + \overline{K}^0)$'s, is positive in 
accordance with the predictions of Ref.~\cite{Olli92} and its strength 
increases with rising $p_t$. 

The dip at $y = 0$, clearly seen in Fig.~\ref{fig4}, looks a bit 
peculiar. However, the signal is not unique. The appearance of similar 
midrapidity dip in the $v_2(y)$ distribution of hadrons, calculated in 
\cite{BlSt00} at full RHIC energy $\sqrt{s} = 200$ AGeV within the 
UrQMD model, is linked to the particle formation time. Our study 
supports this conclusion. Figure~\ref{fig4} shows that the effect is
stronger for high-$p_T$ kaons, which have longer formation time due
to higher momentum. Shorter formation times lead to increase of the
elliptic flow, and the dip vanishes.

\begin{figure}[htp]
\centerline{\epsfysize=16cm \epsfbox{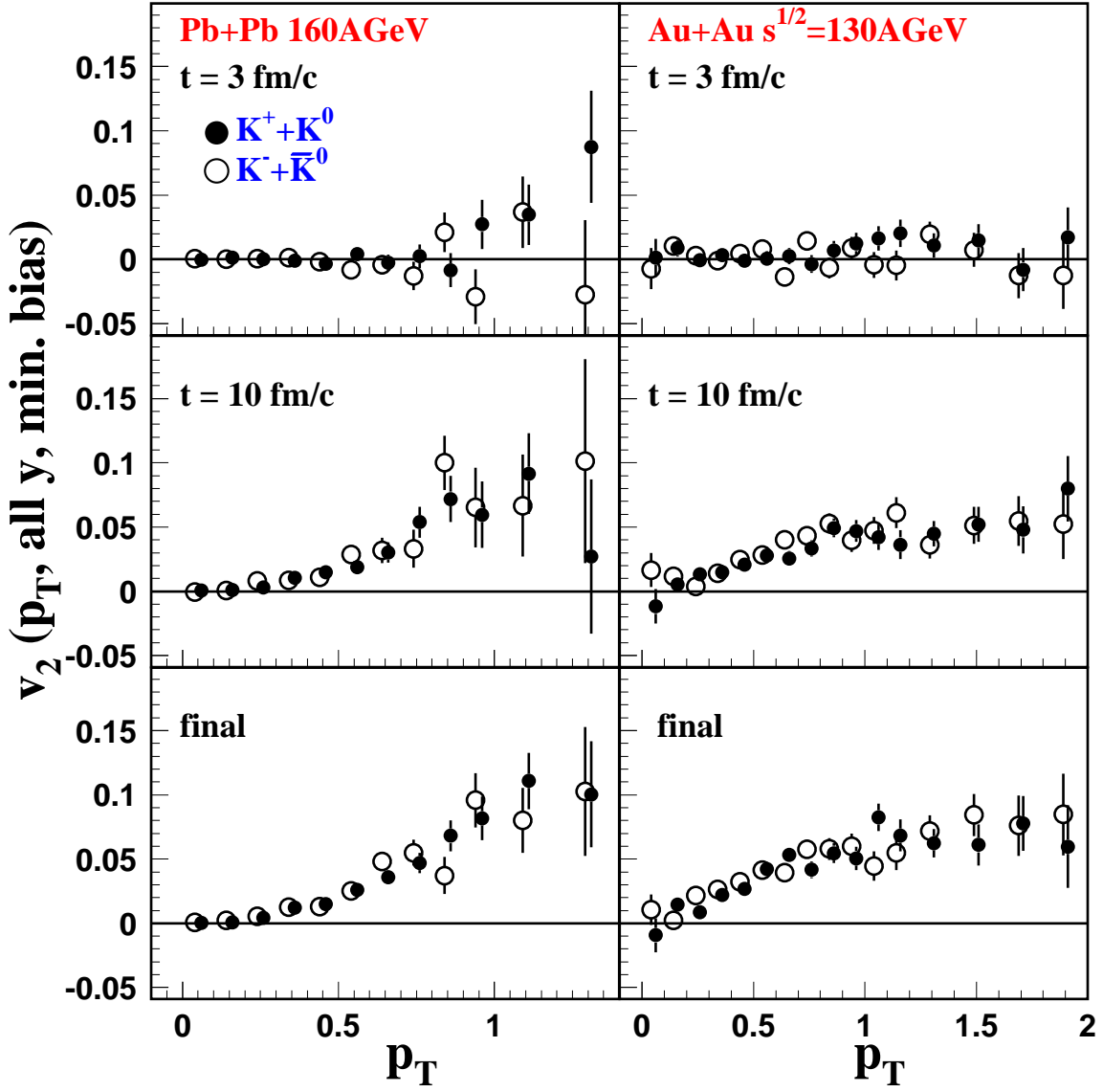}}
\vspace{0.5cm}
\caption{
Transverse momentum dependence of the elliptic flow of kaons (solid
circles) and antikaons (open circles) in minimum bias Pb+Pb collisions
at SPS (left panels) and Au+Au collisions at RHIC (right panels)
at early times $t = 3$ fm/$c$ (upper row), 10 fm/$c$ (middle row),
and freeze-out times (bottom row) of the reaction.
}
\label{fig5}
\end{figure}

The transverse momentum dependence of the elliptic flow in both 
reactions is presented in Fig.~\ref{fig5}. Again, in the very early 
phase of the collision the flow $v_2^{K(\overline{K})} (p_t)$ is 
absent. At time $t = 10$ fm/$c$ the elliptic flow of kaons and 
antikaons seems to be already formed. It is close to zero for kaons 
and antikaons with low transverse momenta, $p_t \leq 0.25$ GeV/$c$, 
and then rises linearly up to $v_2^K (p_t) \approx 10\%$ within the 
interval $0.25 \leq p_t \leq 1.5$ GeV/$c$. At transverse momenta 
higher than 1.5 GeV/$c$ the elliptic flow saturates in accord with 
experimental results \cite{Sne01}. Note, that if the hydrodynamic 
regime would be reached, the excitation function of the elliptic flow 
$v_2^K (p_t)$ should grow further almost linearly above 
$p_t \geq 1.5$ GeV/$c$ also \cite{KSH99}.
 
The delay in the development of the elliptic flow and the saturation
of the flow after a certain $p_t$-limit can be explained as follows.
The bulk production of new particles takes place within the first few 
fm/$c$'s, when parton and hadron collisions are extremely energetic. 
Kaons, as well as other hadrons produced as a result of string 
fragmentation, are emitted isotropically in the azimuthal plane. 
To develop the anisotropic flow these hadrons have to rescatter in a
spatially anisotropic nuclear matter. But because of the uncertainty
principle the only particles allowed to interact immediately (although 
with reduced cross sections) are those containing the valence quarks.
For all other hadrons a proper formation time should pass between the
rescatterings. Due to the finiteness of colliding nuclei, there is a
saturation of number of secondary interactions per hadron with the
transverse momentum higher than certain $p_t$. Therefore, the elliptic 
flow of high-$p_t$ particles also saturates.
Here it is worth to mention that the problem of the formation time of
hadrons produced in a very dense and hot nuclear medium, which is 
formed in heavy ion collisions at energies of RHIC and higher, is 
studied insufficiently. A shorter formation time will reduce the
mean free path of particles, thus increasing the collision rates
and enhancing the elliptic flow. This can be the reason why the
plateau in the excitation function $v_2^{K({\overline K})}(p_t)$,
predicted by the QGSM (see Fig.~\ref{fig5}), is a bit lower compared
to the data. This question should be clarified in nearest future.

\section{Discussion and conclusions}
\label{concl}
In summary, the anisotropic flow of kaons and antikaons is studied 
in lead-lead and gold-gold collisions at SPS and RHIC energies,
respectively, within the microscopic cascade quark-gluon string
model.
The directed flow of kaons $v_1^{K^+ + K^0}(y, \Delta p_t)$ at SPS
energy is close to zero in a broad rapidity range except of the 
projectile and target fragmentation regions, while the directed 
flow of antikaons $v_1^{K^- + \overline{K}^0} (y, \Delta p_t)$ has
a linear antiflow slope. At early times $t \leq 3$ fm/$c$ for both 
particle species the directed flow is oriented in the direction 
of normal flow. The difference in final distributions is 
attributed to the different interaction cross sections with the 
surrounding nuclear matter and to the large absorption cross section 
of antikaons with baryons. In heavy-ion collisions at RHIC energies 
a dense meson-dominated matter with low net baryon density is formed. 
Hence, the directed flow of kaons becomes similar to that of 
antikaons, i.e., in both cases the flow is elongated in antiflow 
direction at $|y| \leq 2$ and in normal flow direction at higher 
rapidity, $|y| \geq 2$. The predicted antiflow of $K$'s in the 
central rapidity range is an order of magnitude stronger compared to 
the signal predicted by the UrQMD model. These are the main results 
of our study. Since the differences between the functions 
$v_1^{K^+ + K^0}(y)$ and $v_1^{K^- + \overline{K}^0} (y)$ in the 
fragmentation regions in different transverse momentum intervals are 
found to be within the statistical error bars, we conclude that the 
formation of fast kaons on leading $u$ and $d$ quarks plays only a 
minor role at RHIC energies.
  
The elliptic flow of both kaons and antikaons in the midrapidity 
range is built up at times between 3 fm/$c$ and 10 fm/$c$, although 
the passing time even for central heavy-ion collisions is 1.44 fm/$c$ 
and 0.12 fm/$c$ at SPS and RHIC energy, respectively. In contrast to 
the midrapidity range, the elliptic flow of kaons and antikaons with 
rapidity $y \geq 1.5$ continues to develop after $t = 10$ fm/$c$ in
the center-of-mass system. The rapidity distribution of the function 
$v_2^{K(\overline{K})}$ is flat over a broad range $|y| \leq 3$ at
RHIC energies. As a function of transverse momentum the elliptic flow
of $K$'s and $\overline{K}$'s rises linearly with increasing $p_t$ 
and saturates at $p_t \geq 1.5$ GeV/$c$.
The plateau at high transverse momenta can be explained by 
saturation of number of secondary interactions experienced by hadrons
with a transverse momentum higher than a certain limiting value.

{\bf Acknowledgments}. 
Fruitful discussions with Yu.~Dokshitzer, J.-Y.~Ollitrault, 
S.~Panitkin, D.~R\"ohrich, D.~Strottman, and N.~Xu are gratefully 
acknowledged.
This work was supported in part by the Bundesministerium f\"ur 
Bildung und Forschung (BMBF) under contract 06T\"U986, and by the
Bergen Computational Physics Laboratory (BCPL) in the framework of the
European Community - Access to Research Infrastructure action of the
Improving Human Potential Programme.

\newpage

\end{document}